\documentclass[prb,reprint,amsmath,amssymb,showpacs,superscriptaddress,floatfix]{revtex4-1}
\usepackage{graphicx,epsfig}% Include figure files
\usepackage{amsmath}% bold math
\usepackage{xcolor}% coloured text
\usepackage[breaklinks=true,colorlinks,citecolor=blue,linkcolor=blue,urlcolor=blue]{hyperref}

\def\be{\begin{equation}}
\def\ee{\end{equation}}

\def\ber{\begin{eqnarray}}
\def\eer{\end{eqnarray}}
\begin{document}

\title{Effective Doping of Monolayer Phosphorene by Surface Adsorption of Atoms for Electronic and Spintronic Applications}

\author{Priyank Rastogi }
\affiliation{Dept. of Electrical Engineering, Indian Institute of Technology Kanpur, Kanpur 208016, India}
\author{Sanjay Kumar}
\affiliation{Dept. of Electrical Engineering, Indian Institute of Technology Kanpur, Kanpur 208016, India}
\author{Somnath Bhowmick}
\affiliation{Dept. of Materials Science and Engineering, Indian Institute of Technology Kanpur, Kanpur 208016, India}
\email{bsomnath@iitk.ac.in}
%\phone{+91-512-2597161}
\author{Amit Agarwal}
\affiliation{Dept. of Physics, Indian Institute of Technology Kanpur, Kanpur 208016, India }
\email{amitag@iitk.ac.in}
\author{Yogesh Singh Chauhan}
\affiliation{Dept. of Electrical Engineering, Indian Institute of Technology Kanpur, Kanpur 208016, India}
\date{\today}
\begin{abstract}

We  study the effect of surface adsorption of 27 different adatoms  on the electronic and magnetic properties  of monolayer black phosphorus using density functional theory. Choosing a few representative elements from each group, ranging from alkali metals (group I) to halogens (group VII), we calculate the band structure, density of states, magnetic moment and effective mass for the energetically most stable location of the adatom on monolayer phosphorene. We predict  that  group I metals (Li, Na, K), and group III adatoms (Al, Ga, In) are effective in enhancing the n-type mobile carrier density, with group III adatoms resulting in lower effective mass of the electrons, and thus higher mobilities. Furthermore we find that the adatoms of transition metals Ti and Fe, produce a finite magnetic moment ($ 1.87$ and $2.31$ $\mu_B$) in monolayer phosphorene, with different band gap and electronic effective masses (and thus mobilities), which approximately differ by a factor of 10 for spin up and spin down electrons opening up the possibility for exploring spintronic applications. 
%Finally we predict group VII halogens (F, Cl and Br) assist in enhancing the p-type mobile carrier density in monolayer phosphorene. 
\end{abstract}

% 
%\pacs{}
%
\maketitle

\section{Introduction}

Thin body materials and devices are being considered as the most promising candidates for 5nm semiconductor technology and beyond by semiconductor industry. \cite{Roadmap}.
%Search for optimal material(s) which allow favorable and aggressive channel length scalability in the post Si nano electronics era, 
This has lead us to actively study various 2D layered crystals such as graphene \cite{graphene1}, silicine \cite{silicene1, silicene2}, germanene \cite{germanane1}, transition metal dichalcogenides (MoS$_2$, MoSe$_2$, WSe$_2$ etc.) \cite{TMD1},  due to their excellent material properties \cite{2D1, 2D2, 2D3, 2D4, review2D} and  reduced short channel effects. Field effect transistors (FET's) based on graphene \cite{Schwierz}, and MoS$_2$ \cite{MoS2_trans}, have already been demonstrated with performance superior to conventional Si based FET's. More recently, layered black phosphorous (dubbed as phosphorene) has joined this promising family of 2D crystals. In contrast to  transition metal dichalcogenides, the carrier (hole) mobility in phosphorene is quite high ($\sim  1000$ cm$^2$/V.s), which in turn gives a I$_{ON}$/I$_{OFF}$ current ratio of $\sim 10^{5}$ at room temperature \cite{Li1, Liu1,APL_Neto,Lu}. 

\begin{figure*} [t]
%\centering
\includegraphics[width=0.75 \linewidth]{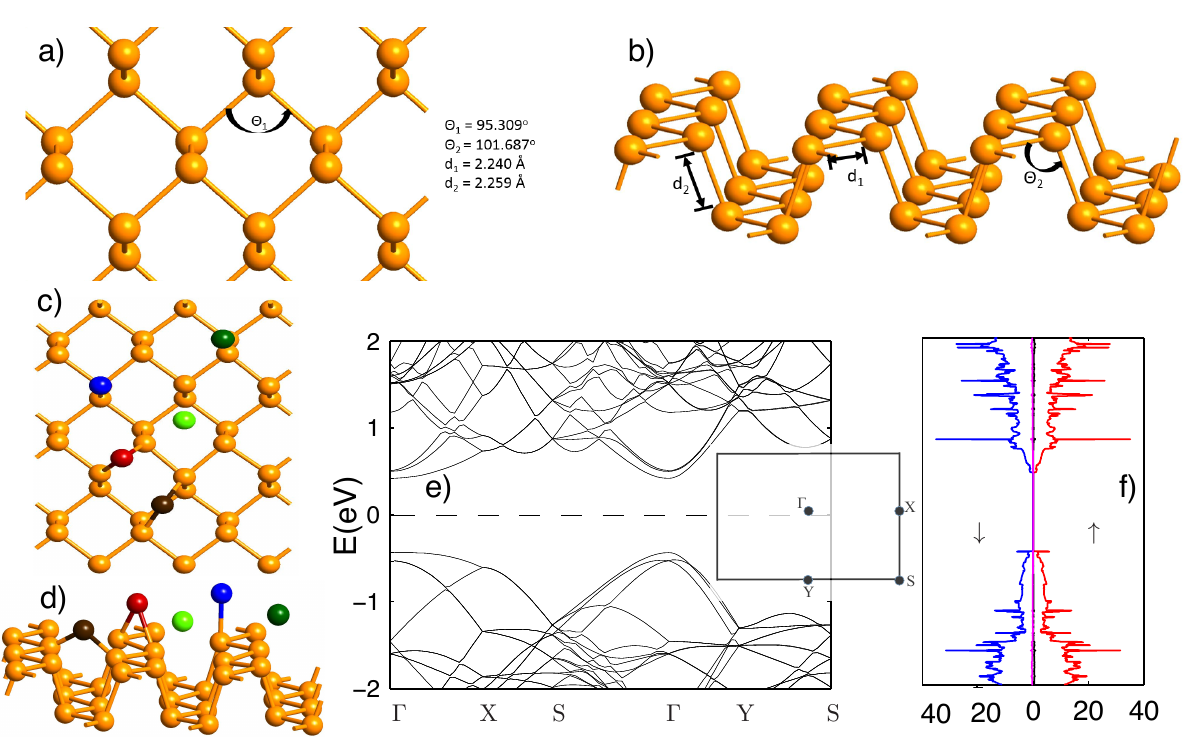}
\caption{(a) Top view and (b) side view of monolayer black phosphorous, (c) top view and (d) side view showing the five different adatom locations: light green at H (hexagon center), dark blue at T$_{\rm u}$ (top of the phosphorous atom in the upper layer), dark green at T$_{\rm l}$ (top of phosphorous atom of lower layer), red at B$_{\rm u}$ (bridge of the upper layer P-P bond) and finally dark brown at B$_{\rm l}$ (bridge of the bottom layer P-P bond). (e) Band structure of intrinsic MBP with dashed line showing the Fermi level, and inset displaying the symmetry points in the Brillouin zone. (f) Spin resolved density of states (DOS) for MBP. Red ($\uparrow$-spin) and blue ($\downarrow$-spin) color lines are for complete DOS of MBP  and black and purple color lines represent projected DOS of P atoms. 
}
\label{fig1}
\end{figure*}

Phosphorene is a layered material with individual layers having a puckered structure and the layers being stacked together by a weak van der Waals interaction, allowing for mechanical exfoliation. In each layer, a phosphorous atom covalently bonds with three other atoms, forming a honeycomb structure. Semiconducting phosphorene has an intrinsic direct bandgap of $\sim 0.31 - 1.0 $ eV (bulk - monolayer), which is inversely proportional to the number of layers \cite{Liu1, Jingsi}. Because of it's direct bandgap at $\Gamma$ point, phosphorene is a potential candidate for applications in the optoelectronic devices\cite{Li1, Liu1, Tran1, Peng, Rodin}. Most of the 2D materials have high electron mobility and display n-type device characteristics, while phosphorene also has a high hole mobility and FET's having both p-type and n-type characteristics have been demonstrated \cite{Li1, Liu1}. Additionally, phosphorene shows highly anisotropic behavior in effective mass and mobility, which can be controlled using appropriate uniaxial or biaxial strain \cite{Fei}. Black phosphorus has also been proposed as an effective anode material for Li ion batteries\cite{C4TA04368E}. Among other advantages, black phosphorous is known for it's chemical inertness and superior transport properties \cite{Tran1,Fei,Peng,Dai,Guo,Tran2,Lu}.
 
For practical use of any semiconductor material in FET devices,  doping is essential to enhance the carrier concentration and to control threshold voltage. Since 2D layered materials have a very high surface area, adatom adsorption can be very effective strategy for carrier doping and band structure engineering. For example, adsorption of hydrogen/halogen, metal and molecules in graphene can change its electronic band structure, transforming  graphene, from metal to semiconductor/insulator \cite{sofo,hubert,chan}. Adatom adsorption in 2D transition metal dichalcogenides creates donor or acceptor levels in between the valence and conduction band of the parent material, leading to electron or hole doping of wide bandgap semiconductors like MoS$_2$ \cite{Ataca, JChang, priyank, priyank1}. Both experimental study and \textit{ab initio} simulation of potassium adsorbtion in MoS$_2$ have shown to dope MoS$_2$ with n-type carriers \cite{Fang,priyank1}. Similar studies have been done  for selected  adatom adsorption in monolayer black phosphorous range of valences, including s and
p valence metals, transition metals, and semiconductors,
hydrogen and oxygen. \cite{C4CP03890H, JPCC_Hu}  Further oxidation of phosphorene has also been studied and may provide more stability and alternate route to synthesis of phosphorene. \cite{PhysRevB.91.085407,C4NR05384B}.

In this article, we use {\it ab-initio} spin polarized density functional theory (DFT) calculations to systematically explore the possibility of band structure engineering and doping possibilities of monolayer black phosphorous (MBP or monolayer phosphorene) via surface adsorption of various adatoms. 
We consider a few representative elements from several group in the periodic table, ranging from alkali metals (group I) to halogens (group VII) and our aim is to provide a comprehensive guideline for doping strategies of MBP via surface adsorption of adatoms. In addition, we  investigate the structural and magnetic properties, and the charge transfer from the adsorbed adatom to MBP  and vice versa. Five different sites are chosen for adatom adsorption, as shown in ~\ref{fig1}(c) and (d). These are named as T$_{\rm u}$ (on top of the  phosphorous atom in upper layer), B$_{\rm u}$ (top of bridge of upper layer),  H (centre of the hexagon), T$_{\rm l}$ (on top of phosphorous atom in lower layer) and finally B$_{\rm l}$ (above the bridge in the lower layer). Our work broadly reaffirms the findings of Ref.~[\onlinecite{C4CP03890H}], and extends the study to include more adatoms such as group III elements (Al, Ga, In)  among others, which are the most effective electron dopant with the lowest effective mass and possibly highest electron mobility. Additionally we also report the effective mass of spin polarized carriers, which is useful for exploring possible usage in electronic and spintronic applications. 

The paper is organized as follows: in Sec.~\ref{Sec2}, we present the parameters and other details of the computational method used. In Sec.~\ref{Sec3}, we briefly review the crystal structure and electronic properties of MBP. This is  followed by systematic presentation of the results and discussion in Sec.~\ref{Sec4} and finally we summarize our findings in Sec.~\ref{Sec5}. 

\begin{figure} [t]
%\centering
\includegraphics[width=0.9 \linewidth]{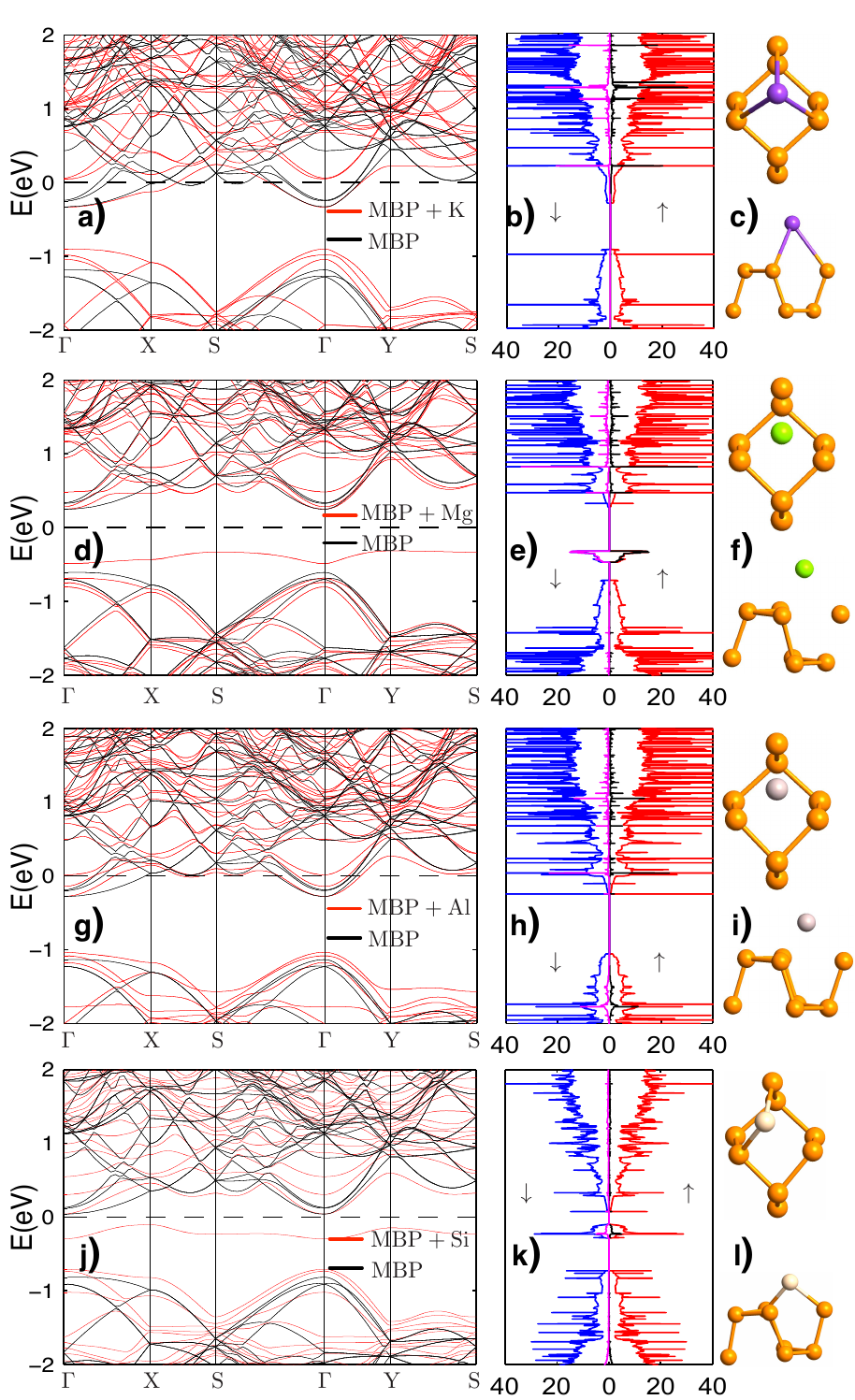}
\caption{ Panels a), d), g) and j) show the band structure, panels b), e), h) and k) display the spin polarized total (red for $\uparrow$-spin and blue for $\downarrow$-spin) and projected DOS (black for $\uparrow$-spin and magenta for $\downarrow$-spin) of MBP with selected adatoms of group I, II, III, and IV, i.e. K, Mg, Al and Si respectively. Panels c), f), i) and l) show the top and side view of the optimized adatom location. 
\label{fig2} }
\end{figure}

\section{Methodology and simulation details}
\label{Sec2}

We use the spin polarized density functional theory (DFT) calculations, as implemented in the ATK package \cite{ATK,ATK1,ATK2} for geometry relaxation, as well as the electronic and magnetic property predictions. The exchange-correlation energy is treated by the generalized gradient approximation (GGA) using the Perdew, Burke, and Ernzerhof (PBE) pseudo potential \cite{PBE}. The Brillouin-zone integrations are performed with Monkhorst-Pack K-points grid with 8x8x1 Brillouin-zone sampling, along with energy cut off of 75 Hartree. We have used a  double zeta polarized basis set for all the calculations performed in this work. We have used a 3x3x1 simple orthorhombic phosphorene super cell (containing 36 phosphorous atoms) with a vacuum region of 10~\AA~ along the $c$-axis, which is sufficient to eliminate interaction among monolayers. 

The charge transfer from the adatom to the phosphorene substrate is calculated using the Mulliken charge transfer method, i.e.,  
by calculating the difference of valence electron charge of the isolated adatom and charge of the adatom in the vicinity of phosphorene;  $\rho*= Z_a- \rho_a$ \cite{Mulliken}. Here $Z_{a}$ indicates the valance electron charge of the isolated adatom with an appropriate sign, for eg. Mg has $Z_a = -2 e$ where $e \approx 1.6 \times 10^{-19}$ eV, and $\rho_a$ is the calculated electron charge on the adatom over the monolayer phosphorene layer. Thus, $\rho^* < 0 $ ($\rho^* > 0 $) implies excess electron charge transferred from (to) the adatom to (from) the phosphorene monolayer.

To establish the stability of each of the reported adatom location on phosphorene monolayer,  we perform ab-initio molecular dynamics simulations using Quantum Espresso (QE) package \cite{QE}.  The MBP-adatom supercell is equilibrated at 400 K temperature and the dynamics of the adatom is monitored for 1 ps. The kinetic energy cutoff of is taken to be 40 Ry. The exchange-correlation energy is treated using PBE pseudopotential and van der Waals correction is included under Grimme-D2  approximation \cite{Grimme}. Ionic temperature is controlled using Anderson thermostat. We find that while all of the adatoms vibrate around their most stable locations (reported in table I), there is no significant shift or dislocation from the most stable location at 400 K.

Additionally we also study the adatom mobility and diffusion kinetics on monolayer phosphorene using the nudge elastic band (NEB) approach which calculates the activation energy barriers encountered by the adatom in moving from one stable location to another. The NEB is fairly accurate approach to determine minimum energy path between initial and final positions \cite{NEB1, NEB2}. For this, Broyden optimization scheme has been employed with eight intermediate images between initial and final positions to find the potential energy barrier between them. We find that for all the twenty seven adatoms considered by us,  the minimum potential energy barrier is 0.6 eV for moving from the most stable location to the next stable location. Since the minimum potential energy barrier of 0.6 eV is  much greater than the available thermal energy at room temperature (0.026 eV),  this implies the  stability of all considered adatom against diffusion from their stable location.

{ Finally for further confirmation of the binding energy we redo the spin polarized virtual crystal relaxation calculations in Quantum Espresso package \cite{QE} taking van der Waals forces into account. This was necessitated on account of the discrepancy in some of our ATK reported binding energies, with those reported in Refs.~[\onlinecite{C4CP03890H, JPCC_Hu}]. For the binding energy verification in QE we use the projector augmented wave (PAW) pseudopotentials within the GGA-PBE approximation and the modified results of both the calculations are shown in Table I.}

\section{Phosphorene: crystal structure and electronic properties} 
\label{Sec3}
Black phosphorus (BP) is reported to be the most stable form among all the allotropes of phosphorus (for example red and white phosphorous among others) under normal conditions. BP has an orthorhombic crystal structure (space group: Cmca), having eight phosphorous atoms in the unit cell \cite{Crichton}. Multiple layers are stacked together to form BP, although unlike graphite the constituent layers are not flat, but puckered in nature. Based on tight binding calculations BP was found to be a direct bandgap semiconductor with an energy gap of $\approx$ 0.3 eV at the Z point \cite{Takao}, which was later confirmed via DFT calculations by \citeauthor{Jingsi}\cite{Jingsi} The authors also calculated the band structure of MBP and found it to be a direct bandgap semiconductor with an energy gap  of $\sim 1$ eV  at the $\Gamma$ point. Our calculation yields a very similar result for pristine MBP [see \ref{fig1}(e)-(f)] and this also validates the parameters chosen for simulation in this article, as described in Section~\ref{Sec2}.

\begin{figure} [t]
%\centering
\includegraphics[width=0.9 \linewidth]{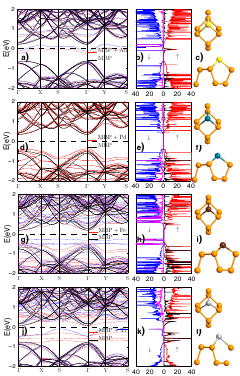}
\caption{ Panels a), d), g) and j) show the band structure, panels b), e), h) and k) display the spin polarized total (red for $\uparrow$-spin and blue for $\downarrow$-spin) and projected DOS (black for $\uparrow$-spin and magenta for $\downarrow$-spin) of MBP with selected metallic adatoms, i.e. Au, Pd, Fe and Ti respectively. Panels c), f), i) and l) show the top and side view of the optimized adatom location.
\label{fig3} }
\end{figure}

\begin{figure} [t]
%\centering
\includegraphics[width=0.8 \linewidth]{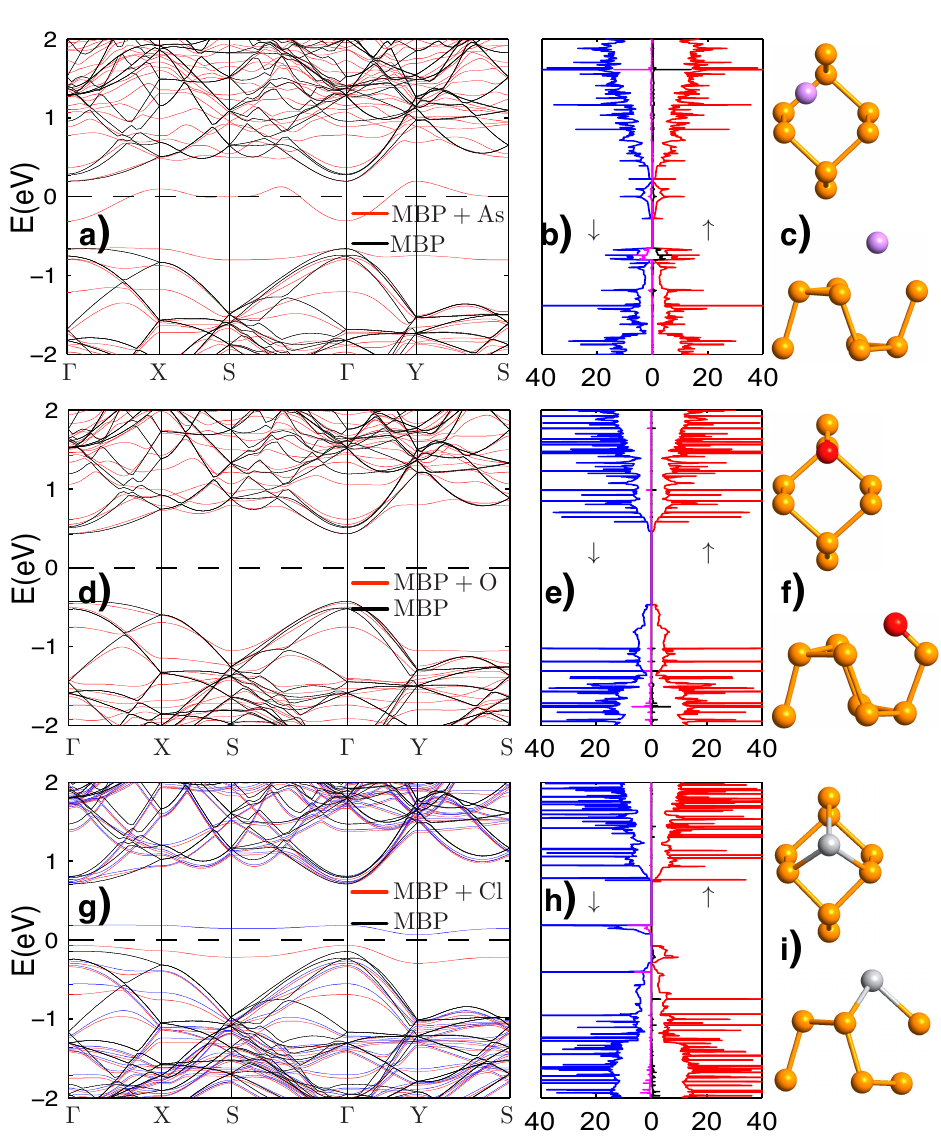}
\caption{ Panels a), d), and g) show the band structure, panels b), e), and h)  display the spin polarized total (red for $\uparrow$-spin and blue for $\downarrow$-spin) and projected DOS (black for $\uparrow$-spin and magenta for $\downarrow$-spin) of MBP with selected adatoms of group V, VI, and VII, i.e. As, O, and Cl respectively. Panels c), f), and i) show the top and side view of the optimized adatom location.
\label{fig4} }
\end{figure}

\section{Results and discussions}
\label{Sec4}

As shown in \ref{fig1}, we have considered five different adatom locations -- H, T$_{\rm u}$, T$_{\rm b}$,  B$_{\rm t}$ and  B$_{\rm l}$ for the adatom. We start by initially placing all the adatoms studied at these five locations, and let the structure relax. The most stable configuration is determined based on maximum binding energy and is chosen for the purpose of presenting the electronic band-structure data. The binding energy is calculated according to the relation:
\be
\label{eq1}
E_{\rm binding}= E_{\rm MBP} + E_{\rm adatom}  -  E_{\rm MBP + adatom}~,
\ee
where $E_{\rm MBP}$ is the total energy of the super cell of pristine MBP, $E_{\rm adatom}$ is the energy of the isolated adatom in the ground state calculated for the same super cell with the same parameters, and   $E_{\rm MBP+adatom}$ is the total energy of the combined phosphorene--adatom system after structural relaxation. In \ref{T1}, we have reported the maximum binding energy value among the five different adatom locations (H, T$_{\rm u}$, T$_{\rm l}$, B$_{\rm u}$ and  B$_{\rm l}$) and presented the corresponding electronic band structure, DOS and adatom projected density of states (PDOS) results in \ref{fig2}, \ref{fig3} and \ref{fig4}. Note that in these figures the dashed line displays the Fermi level of the MBP-adatom system and the band structure of MBP (black lines) has been shifted so as to align the conduction bands of pristine MBP and the MBP-adatom system.

Let us begin by discussing the case of alkali and alkaline earth metals. As shown in \ref{T1}, all of them prefer H site among the five possible locations and the vertical distance ($d$) from the upper phosphorous layer  varies depending on the size of the adatom. Smaller atoms have higher binding energies, which implies stronger binding to the underlying MBP layer. In case of alkali metals (group I), adatom adsorption leads to n-type carrier doping, because of which Fermi level shifts to the conduction band of MBP [see the case of K adsorption in \ref{fig2}(a)]. For each of the three group I adatoms studied here, MBP turns out to be electron doped; as expected because of highly electropositive nature of alkali metal atoms. Since the electronic states of K have negligible contribution to the total density of states (DOS) near the CBM [see \ref{fig2}(b)], the K adatom (and other alkali metals) preserves the original electronic structure of phosphorene, but shifts the Fermi level to the CBM by donating free mobile electrons, and hence enhances the electrical conductivity significantly.
 In case of group II, adatom adsorption yields significantly different effect. As shown in \ref{fig2}(d), after Mg adsorption a filled energy band is present in the band gap of MBP. Electronic states corresponding to this particular band are localized near the adatom, which is clearly visible from the PDOS plot shown in \ref{fig2}(e). Thus, the only way these electrons can take part in electronic conduction is via excitation to the conduction band of MBP. However, since the band is located $\sim 0.7$ eV away from the CBM, group II adatoms (Mg, Ca studied here) are not going to be effective as electron donors. The PDOS plots [see \ref{fig2}(b) and (e)] reveal that the up and down spin states have mirror symmetry, which implies equal number of up and down spin electrons and accordingly all the adsorbed atoms of group I and II are non-magnetic.

\begin{table*}
\caption{Adatom, its preferred adsorption site [see \ref{fig1} c) -d)], its binding energy  (calculated using Eq.~\ref{eq1}), vertical distance of the adatom $d$ from the upper phosphorous layer \footnote{Note that $d$ is not the shortest distance from the nearest phosphorous atom}, magnetic moment per adatom (in the units of Bohr magneton, $\mu_B$) in a $3 \times 3 \times 1$ supercell, electronic charge transfer and the electronic effective mass (in the units of electron rest mass, m$_e$) calculated at the $\Gamma$ point in the $\Gamma-$X direction. \label{T1}}
\begin{tabular}{c c c c c c c  c c c}
\hline \hline
& & Adsorption & ~~~Binding energy\footnote{ Calculated using: (i) Quantum Espresso package --- PAW-PBE pseudopotential and vdW correction (ii) QuantumWise ATK --- norm conseving basis set with PBE and no vdW correction}~~~~ & Height & Magnetic moment & Charge & Effective\\
Group & Adatom & Site & (eV) & d (nm) & on adatom ($\mu_B$)& transfer (e) & mass ($m_{\rm e}$)\\
& & & (i)  ,  (ii) & & & &  \\
%& & & & (nm) & on adatom  &  (e) & $m_b$/$m_e$\\
%& & & &      & ($\mu_B$)      &       &     ~     \\

\hline 
I$^{\rm st}$ & Li & H & 1.922 , 2.545 & 0.146 & 0 & -0.024 & 1.26 \\
             & Na & H & 1.677 , 1.554 & 0.200 & 0 & -0.176 & 1.24 \\
             &  K & H & 2.01  , 1.576 & 0.258 & 0 & -0.348 & 1.27 \\
\hline
II$^{\rm nd}$  
             & Ca  & H & 2.193 , 2.513 & 0.201 & 0 & -0.056 & 0.65\\
	     & Mg  & H & 0.944 , 1.088 & 0.197 & 0 & -0.170 & 0.53\\			
\hline   
III$^{\rm rd}$  & Al   & H & 2.370 , 3.186 & 0.161 & 0 & 0.380 &  0.37\\
                & Ga   & H & 2.071 , 2.901 & 0.172 & 0 & 0.448  & 0.43\\
                & In   & H & 1.970 , 3.124 & 0.182 & 0 & 0.36 & 0.43\\
\hline
IV$^{\rm th}$ & C  & H            & 5.578 , 6.330  & 0.045 & 0 &-0.01  & 0.576 \\
              & Si  & B$_{\rm l}$ & 3.130 , 4.284 & 0.129 & 0 & 0.274 & 0.337  \\ 
              & Ge  & B$_{\rm l}$ & 2.658 , 3.391 & 0.148 & 0 & 0.14  & 0.328  \\ 
\hline
V$^{\rm th}$  & N   & B$_{\rm l}$ & 3.270 , 4.822 & 0.065 & 0.056 & -0.028 & 0.746($\uparrow$),0.763($\downarrow$)\\	
              & As  & B$_{\rm l}$ & 2.015 , 2.408 & 0.143 & 0 & 0.036 & 0.362\\
\hline
VI$^{\rm th}$ & O   & T$_{\rm l}$  & 5.627 , 5.950 & 0.092 & 0 & 0.136 & 0.696 \\ 		
	      & S   & T$_{\rm l}$  & 3.368 , 3.894 & 0.153 & 0 & 0.104 & 0.650 \\	
\hline
VII$^{\rm th}$ & F  &  T$_{\rm u}$ & 3.428 , 4.005 & 0.181 & 0  & 0.206 & 0.699($\uparrow$),0.728($\downarrow$)    \\
               & Cl &  T$_{\rm u}$ & 1.829 , 2.442 & 0.218 & 0.14  &  0.124 & 2.157($\uparrow$),1.881($\downarrow$)\\
               & Br & T$_{\rm u}$  & 1.428 , 2.061 & 0.231 & 0.155 & 0.149   & 1.718($\uparrow$),1.465($\downarrow$)\\
 \hline
Nobel        & Cu & H & 2.485 , 3.556 & 0.113 & 0  & 0.214 & 0.836 \\
 Metals      & Ag & H & 1.359 , 2.270 & 0.146 & 0  & 0.046 & 0.494 \\
             & Au & H & 2.301 , 2.625 & 0.150 & 0.242 & 0.064 & 0.502($\uparrow$),0.429($\downarrow$)\\
 \hline
Transition   & Ti & H & 3.858 , 4.519 & 0.150 & 1.874 & -0.056 & 1.471($\uparrow$),0.158($\downarrow$) \\
             & Fe & H & 4.604 , 4.293 & 0.112 & 2.316 & -0.006 & 2.825($\uparrow$),0.324($\downarrow$)\\
metals       & Ni & H & 4.330 , 5.524  & 0.109 & 0 & -0.042 & 0.765\\
             & Pd & H & 3.567 , 4.547 & 0.133 & 0 & 0.522 & 0.602\\
	     & Pt & H & 5.236 ,  6.099 & 0.121 & 0 & 0.208 & 1.856 \\
             & Sn & H & 2.252 , 2.932 & 0.164 & 0 & 0.092 & 0.305\\
\hline
\hline 
\end{tabular}
\end{table*}

Next, we consider the group III and IV adatoms. In case of group III, all the adatoms (Al, Ga, In) are located on the H site and they have similar binding energies [see \ref{T1}]. Among the group IV adatoms C prefers the H site, while Si and Ge are more stable on the B$_{\rm l}$ site and the binding energy is strongly dependent on the size of the adatom [see \ref{T1}]. Adatom-phosphorene distance $d$ depends on the size of the corresponding adatom, as expected [see \ref{T1}]. In case of group III, as a result of adatom adsoprtion, the Fermi level has shifted to the conduction band of MBP, making it a n-type semiconductor [see \ref{fig2}(g)]. On the other hand adsorption of group IV adatoms leads to the formation of fully occupied midgap energy levels, located approximately in the middle of VBM and CBM [see \ref{fig2}(j)]. Thus, while group III adatoms are very effective for n-type carrier doping in MBP, group IV adatoms are not suitable at all for this purpose.  We have not found magnetic moment on any of the adatoms of group III and IV, which is corroborated by DOS plots of spin up and spin down states, as illustrated in ~\ref{fig2}(h) and (k).  We emphasize here that only considering the charge transfer to determine weather an adatom adsorption leads to electron or hole doping can be misleading, as in the case of group III  elements [see \ref{T1}]. In case of group III adatoms, the changes in the band structure lead to a suppression of electronic states in the valance band, and an enhancement of DOS in the conduction band [see \ref{fig1}h]. Also note that group III adatoms in MBP, leads to a lower effective mass of the electrons ($ \sim 0.4 m_e$), which implies higher mobilities and consequently excellent transport properties.
   
Now we consider the case of the  transition metals. According to the binding energy values reported in \ref{T1}, all of them get adsorbed preferably on the H site. Among them, only Au, Ti and Fe are found to be magnetic [see \ref{T1}], which is corroborated by up and down spin DOS, shifted with respect to each other along the energy axis [see \ref{fig3}(b), (h) and (k)]. Interestingly, magnitude of the bandgap  for up and down spin electrons are different, which can be exploited for spintronic applications. Also note that the effective electron mass for the up and down spin electrons in Fe and Ti, differ  by a factor of 10 (approximately), which will also be reflected in their respective mobilities. Each one of the adatoms gives rise to the midgap energy levels [note that the Fermi energy lies in between VBM and CBM in \ref{fig3}(a), (d), (g) and (j)] and thus they are not effective either as n-type or as p-type dopant in MBP.  We note however, that these trap states can speed up the carrier recombination process and can be useful for devices which require fast switch off times \cite{Streetman}. 

Finally we describe the effect of group V, VI and VII adatoms on electronic band structure of MBP. Based on the binding energy values shown in \ref{T1}, group V, VI and VII adatoms prefer B$_{\rm l}$, T$_{\rm l}$ and T$_{\rm u}$ site, respectively. As expected the binding energy, as well as the value of $d$ depends on the size of the adatom. As reported in \ref{T1}, smaller adatoms are closer to the MBP layer and bind more strongly than compared to the larger ones. In case of group V, adatom adsorption does not dope the underlying MBP layer with n-type carriers directly [see the case of As in \ref{fig4}(a)]. However, the energy gap (indirect) between the midgap state and the CBM is very small, such that these electrons can be excited to the conduction band of MBP to increase the n-type carrier density. On the other hand, the bandgap remains unchanged by the adsorption of group VI adatoms [see the case of O in \ref{fig4}(d)] and thus they are not effective for carrier doping in MBP. Among the group VII adatoms, while F is found to be non-magnetic, a small magnetic moment ($\approx$ 0.15 $\mu_B$ per adatom) is observed in case of Cl and Br. This is confirmed by the asymmetry of the DOS plots of up and down spin states in case of Cl adsorption, illustrated in \ref{fig4}(g). Note that, the bandgap of up and down spin electrons have different magnitude, which can be useful for spintronic device applications. Moreover for group VII adatoms, the Fermi energy lies close to the valance band, which may assist in making MBP hole-doped. 

{ Before concluding let us compare the reported binding (or adsorption) energy in previous works \cite{C4CP03890H, JPCC_Hu} and in this work. Kulish et al. in Ref.~[\onlinecite{C4CP03890H}], did their calculations using QE, considering PAW method with PBE as the exchange correlation functional, while ignoring dipole- as well as  van der Waals- corrections. 
Hu et al. in Ref.~[\onlinecite{JPCC_Hu}], used VASP, with PAW-PBE functional, included dipole corrections and ignored van der Waals corrections. In this paper,  we have used two different methods for calculating the binding energies: (1) using norm conserving pseudo potential with PBE functional in ATK, ignoring van der Waals forces and (2) using PAW-PBE functional in QE, taking van der Waals forces into account, respectively. We find that the binding energy depends significantly on type of pseudo potential used and weather or not the van der Waals correction is included. Further since our QE calculations use the same basis set and exchange correlation functional as Ref.~[\onlinecite{C4CP03890H, JPCC_Hu}], the QE calculated binding energies are also similar. } 

\section{Conclusion}
\label{Sec5}
To summarize, in this article we have systematically explored the effect of surface adsorption of 27 adatoms from different groups of the periodic table, ranging from alkali metals (group I) to halogens (group VII), on electronic, structural and magnetic properties of MBP. % using spin polarized density functional theory. 
We find that among the adatoms considered, group I alkali metals and group III elements make MBP n-type, with the  MBP-group III adatom system having lower electronic effective mass ($\sim 0.4 m_e$) which should result in better mobility and transport characteristics. Group II, group IV, and transition metals give rise to mid gap states, and are not effective as dopants. 
However, Au, Ti and Fe adatoms induce spin-polarization  with  magnetic moments of $0.24 $, $ 1.87$ and $2.31$ $\mu_B$, respectively, in a $3\times3\times1$ super cell. Moreover in the case of Fe and Ti adatoms the effective mass of the up and down electrons differ approximately by a factor of 10,  are likely to have different mobilities and thus they may be very useful for spintronic applications.
Amongst the adatoms of group V, VI, and VII, while none of them makes MBP either n-type or p-type directly, for the case of group VII elements  the Fermi level lies close to VBM and it may be useful for increasing hole concentration. Finally we have also checked the stability of all the adatoms at their most stable location, against temperature fluctuations by doing a MD simulations, and against surface diffusion by calculating the activation energy barriers for adatoms moving to other locations.

%Our results predict various doping strategies for MBP via surface adsorption of adatoms, which may be tested experimentally and will be useful for exploring the applications of phosphorene in diverse fields such as energy storage, catalysis, sensors, nano-electronics and so on.

\section*{Acknowledgements}  
The authors acknowledge funding by Ramanujam fellowship research grant, DST fast-track scheme for young scientists, IBM faculty award, Semiconductor Research corporation, Faculty Initiation grants by Indian Institute of Technology Kanpur (India) and from INSPIRE faculty fellowship by DST.

%\bibliography{Phosphorene_adatom_v4_PRB}

%merlin.mbs apsrev4-1.bst 2010-07-25 4.21a (PWD, AO, DPC) hacked
%Control: key (0)
%Control: author (72) initials jnrlst
%Control: editor formatted (1) identically to author
%Control: production of article title (-1) disabled
%Control: page (0) single
%Control: year (1) truncated
%Control: production of eprint (0) enabled
%

\end{document}